\documentclass[sigconf]{acmart}

\def\BibTeX{{\rm B\kern-.05em{\sc i\kern-.025em b}\kern-.08emT\kern-.1667em\lower.7ex\hbox{E}\kern-.125emX}}

\usepackage[nolist]{acronym}
\usepackage{nicefrac}
\usepackage{siunitx}
\usepackage{array,framed}
\usepackage{booktabs}
\usepackage{
  color,
  float,
  epsfig,
  wrapfig,
  graphics,
  graphicx,
  subcaption
}

\usepackage{setspace}
\usepackage{latexsym,fancyhdr,url}
\usepackage{enumerate}
\usepackage{algorithm2e}
\usepackage{algpseudocode}
\usepackage{graphics}
\usepackage{xparse} 
\usepackage{xspace}
\usepackage{multirow}
\usepackage{csvsimple}
\usepackage{balance}
\usepackage{subcaption}
\usepackage{booktabs}
\usepackage{soul}

\usepackage[bottom]{footmisc}
\usepackage{tabularx}
\usepackage{amsfonts}
\usepackage{adjustbox}
\usepackage[english]{babel}
\usepackage{colortbl}
\usepackage{dirtree}
\usepackage{enumerate}
\usepackage{hyperref}  
\usepackage[utf8]{inputenc}
\usepackage{csquotes}
\usepackage{lscape}
\usepackage{longfigure}
\usepackage{longtable}
\usepackage{multirow}
\usepackage{multicol}
\usepackage{subcaption}
\usepackage{xspace}
\newcommand{\mypar}[1]{\smallskip\noindent\textbf{#1.}\xspace}
\newcommand{\mypara}[1]{\smallskip\noindent\textbf{#1:}\xspace}

\usepackage{listings}
\usepackage{graphicx}
\usepackage{pgf-umlsd}
\usepackage{afterpage}
\usepackage{amsmath}
\usepackage{rotating}   
\usepackage{fancyhdr}

\usepackage{tikz-timing}[2014/10/29]
\usetikztiminglibrary[rising arrows]{clockarrows}
\usepackage{forest,array}
\usepackage{textcomp}
\useforestlibrary{edges}
\usetikzlibrary{shadows}
\newcolumntype{C}[1]{>{\centering}p{#1}}

\newcommand{\system}{CANflict }
\newcommand{\systemns}{CANflict}

\newcommand{\master}{{primary}}
\newcommand{\slave}{{secondary}}

\newcommand{\MISOns}{{CIPO}}
\newcommand{\MOSIns}{{COPI}}

\newcommand{\MISO}{\MISOns \xspace}
\newcommand{\MOSI}{\MOSIns \xspace}

\newcommand{\masters}{\master \xspace}
\newcommand{\slaves}{\slave \xspace}

\makeatletter
\def\BState{\State\hskip-\ALG@thistlm}
\makeatother

\NewDocumentCommand{\busref}{som}{\texttt{%
#3%
\IfValueTF{#2}{[#2]}{}%
\IfBooleanTF{#1}{\#}{}%
}}

\definecolor{mygreen}{HTML}{D5E8D4}
\definecolor{myred}{HTML}{F8CECC}
\definecolor{myorange}{HTML}{FAD7AC}
\definecolor{myblue}{HTML}{DAE8FC}
\definecolor{mydarkgreen}{HTML}{82B366}
\definecolor{mydarkred}{HTML}{B85450}
\definecolor{mydarkorange}{HTML}{B46504}
\definecolor{mydarkblue}{HTML}{6C8EBF}

\tikzset{
  basic/.style = {
    draw, text width=4cm, thin, align=center,
    drop shadow, font=\fontsize{6pt}{12pt}\selectfont,
    fill=myblue, draw=mydarkblue
    },
  root/.style = {
    basic, rounded corners=2pt,
    text width=1.5cm,
   fill=mygreen, draw=mydarkgreen},
  level 2/.style = {
    basic, rounded corners=6pt, align=left,
    fill=myred, draw=mydarkred, text width=5em},
  level 3/.style = {
    basic,
    align=left,
    fill=myorange, draw=mydarkorange, text width=5em},
    level 4/.style = {
        basic,
        align=left,
        fill=myred, draw=mydarkred, text width=5em}
}

\def\acmBooktitle#1{\gdef\@acmBooktitle{#1}}
\acmBooktitle{Proceedings of \acmConference@name
       \ifx\acmConference@name\acmConference@shortname\else
         \ (\acmConference@shortname)\fi}

\begin{acronym}
	{\small
	\acro{SPI}{Serial Peripheral Interface}
	\acro{CPS}{Cyber-physical System}
	\acro{CAN}{Controller Area Network}
	\acro{CAN-FD}{Controller Area Network with Flexible Data-rate}
	\acro{DSRC}{Dedicated Short Range Communications}
	\acro{ECU}{Electronic Control Unit}
	\acro{TCU}{Telematic Control Unit}
	\acro{MCU}{Microcontroller Unit}
	\acro{GPIO}{General Purpose Input/Output}
	\acro{IRQ}{Interrupt Request}
	\acro{OBD-II}{On-Board Diagnostic}
	\acro{OBD}{On-Board Diagnostic}
	\acro{IoT}{Internet of things}
	\acro{OEM}{Original Equipment Manufacturer}
	\acro{MOST}{Media Oriented System Transport}
	\acro{ADAS}{Advanced Driver Assistance System}
	\acro{ABS}{Antilock Braking System}
	\acro{LIN}{Local Interconnect Network}
	\acro{TPMS}{Tire Pressure Monitoring System}
	\acro{UDS}{Unified Diagnostic System}
	\acro{IDS}{Intrusion Detection System}
	\acro{ADS}{Anomaly Detection System}
	\acro{LSTM}{Long Short-Term Memory}
	\acro{OTA}{Over The Air}
	\acro{RNN}{Recurrent Neural Network}
	\acro{HMM}{Hidden Markov Model}
	\acro{ACK}{Acknowledgment}
	\acro{CRC}{Cyclic Redundancy Check}
	\acro{IFS}{Interframe Space}
	\acro{TEC}{Transmit Error Count}
	\acro{REC}{Receive Error Count}
	\acro{LTE}{Long Term Evolution}
	\acro{DoS}{Denial of Service}
	\acro{BC}{Bit Counter}
	\acro{PC}{Polarity Counter}
	\acro{SoF}{Start of Frame}
	\acro{DLC}{Data-Length Code}
	\acro{IDE}{Identifier Extension}
	\acro{RPM}{Repetitions Per Minute}
	\acro{RTR}{Remote Transmission Request}
	\acro{EoF}{End of Frame}
	\acro{HSM}{Hardware Security Module}
	\acro{IF}{Intermission Field}
	\acro{DL}{Deep Learning}
	\acro{ML}{Machine Learning}
	\acro{OCSVM}{One-Class SVM}
	\acro{CAN-H}{CAN High}
	\acro{CAN-L}{CAN Low}
	\acro{IRS}{Intrusion Reaction System}
	\acro{CRF}{Conditional Random Field}
	\acro{ROC}{Receiver Operating Characteristic Curve}
	\acro{AUC}{Area Under the Curve}
	\acro{MAC}{Message Authentication Code}
	\acro{GRU}{Gated Recurrent Unit}
	\acro{V2X}{Vehicle to Everything}
	\acro{FMS}{Fleet Management System}
	\acro{TCS}{Traction Control System}
	\acro{DBC}{CAN Data Base}
	\acro{FPR}{False Positive Rate}
	\acro{VAR}{Vector Auto Regression}
    \acro{NN}{Neural Network}
	\acro{ELU}{Exponential Linear Unit}
	\acro{MSE}{Mean Squared Error}
	\acro{TPR}{True Positive Ration}
	\acro{DR}{Detection Rate}
	\acro{MCC}{Matthews Correlation Coefficient}
	\acro{TTP}{Testing Time per Packet}
	\acro{AR}{Auto Regressive}
	\acro{FP}{False Positive}
	\acro{TEE}{Trusted Execution Environment}
	\acro{FCD}{Floating Car Data}
	\acro{GPS}{Global Positioning System}
	\acro{GNSS}{Global Navigation Satellite System}
	\acro{P2P}{Peer-to-Peer}
	\acro{PoW}{Proof-of-Work}
	\acro{PoS}{Proof-of-Stake}
	\acro{PBFT}{Practical Byzantine Fault Tolerance}
	\acro{FCC}{Federal Communications Commission}
	\acro{PRNG}{Pseudorandom Number Generator}
	\acro{MoST}{Monaco SUMO Traffic}
	\acro{GSM}{Global System for Mobile Communications}
	\acro{PoI}{Proof-of-Importance}
	\acro{PoB}{Proof-of-Believability}
	\acro{ITS}{Intelligent Transportation System}
	\acro{IVI}{In-Vehicle Infotainment}
	\acro{C-V2X}{Cellular V2X}
	\acro{RSU}{Road Side Unit}
	\acro{PKI}{Public Key Infrastructure}
	\acro{CDAL}{Controller Data Abstraction Layer}
	\acro{EMB}{Electro-Mechanical Braking system}
	\acro{PID}{Proportional Integral Derivative}
	\acro{DNN}{Deep Neural Network}
	\acro{UART}{Universal Asynchronous Receiver Transmitter}
	\acro{I2C}{Inter-Integrated Circuit}
	\acro{ADC}{Analog-to-Digital Converter}
}
\end{acronym}

\usepackage{
  tikz,
  pgfplots,
  pgfplotstable
}
\usepackage{hyperref}

\usetikzlibrary{
  shapes.geometric,
  arrows,
  external,
  pgfplots.groupplots,
  matrix
}

\pgfplotsset{compat=1.9}

\usepackage{mathtools}

\DeclareMathAlphabet{\mathcal}{OMS}{cmsy}{m}{n}

\DeclareGraphicsExtensions{%
    .png,.PNG,%
    .pdf,.PDF,%
    .jpg,.mps,.jpeg,.jbig2,.jb2,.JPG,.JPEG,.JBIG2,.JB2}

\begin{document}

\copyrightyear{2022} 
\acmYear{2022} 
\setcopyright{acmcopyright}\acmConference[CCS '22]{Proceedings of the 2022 ACM SIGSAC Conference on Computer and Communications Security}{November 7--11, 2022}{Los Angeles, CA, USA}
\acmBooktitle{Proceedings of the 2022 ACM SIGSAC Conference on Computer and Communications Security (CCS '22), November 7--11, 2022, Los Angeles, CA, USA}
\acmPrice{15.00}
\acmDOI{10.1145/3548606.3560618}
\acmISBN{978-1-4503-9450-5/22/11}

\fancyhead{}

\def\thetitle{\systemns: Exploiting Peripheral Conflicts for Data-Link Layer Attacks on Automotive Networks}

\title{\thetitle}
 
\author{Alvise de Faveri Tron}
\email{a.de.faveri.tron@vu.nl}
\affiliation{%
  \institution{Politecnico di Milano}
  \city{Milan}
  \country{Italy}
 }
\affiliation{%
  \institution{Vrije Universiteit Amsterdam}
  \city{Amsterdam}
  \country{Netherlands}
}

\author{Stefano Longari}
\email{stefano.longari@polimi.it}
\affiliation{%
  \institution{Politecnico di Milano}
  \city{Milan}
  \country{Italy}
 }
\author{Michele Carminati}
\email{michele.carminati@polimi.it}
\affiliation{%
  \institution{Politecnico di Milano}
  \city{Milan}
  \country{Italy}
 }
 \author{Mario Polino}
\email{mario.polino@polimi.it}
\affiliation{%
  \institution{Politecnico di Milano}
  \city{Milan}
  \country{Italy}
 }
 \author{Stefano Zanero}
\email{stefano.zanero@polimi.it}
\affiliation{%
  \institution{Politecnico di Milano}
  \city{Milan}
  \country{Italy}
 }
\date{}

%

\begin{abstract}
Current research in the automotive domain has proven the limitations of the \ac{CAN} protocol from a security standpoint. Application-layer attacks, which involve the creation of malicious packets, are deemed feasible from remote but can be easily detected by modern \acp{IDS}. On the other hand, more recent link-layer attacks are stealthier and possibly more disruptive but require physical access to the bus.
In this paper, we present \systemns, a software-only approach that allows reliable manipulation of the CAN bus at the data link layer from an unmodified microcontroller, overcoming the limitations of state-of-the-art works. We demonstrate that it is possible to deploy stealthy \ac{CAN} link-layer attacks from a remotely compromised \acs{ECU}, targeting another ECU on the same CAN network.
To do this, we exploit the presence of \textit{pin conflicts} between microcontroller peripherals to craft \textit{polyglot frames}, which allows an attacker to control the CAN traffic at the bit level and bypass the protocol's rules. 
We experimentally demonstrate the effectiveness of our approach on high-, mid-, and low-end microcontrollers, and we provide the ground for future research by releasing an extensible tool that can be used to implement our approach on different platforms and to build CAN countermeasures at the data link layer.
\end{abstract}

\maketitle

\section{Introduction}
\label{sec:introduction}

Nowadays, vehicles are equipped with an enormous amount of electronic devices~\cite{sensors}, which can include WiFi access points, Bluetooth modules, cellular communication modules, gateways, telemetry systems, and dozens of \acfp{ECU}~\cite{iov}. A modern vehicle, even if not fully-featured, typically has well over 100 \acp{ECU}, with an estimated 7000 signals to transmit internally~\cite{SoftwareEatingCar,longari2019secure}. To coordinate communication among \acp{ECU}, in-vehicle networks employ several kinds of bus protocols. 
The most prevalent and de-facto standard of such protocols is \ac{CAN}.
Developed in the 1980s, the \ac{CAN} protocol was primarily designed for reliable and fast communications in noisy environments, without much consideration for security aspects. The lack of encryption, authentication, and integrity checking makes CAN bus networks vulnerable to different attacks. Such attacks have first been proven possible through on-board attack surfaces~\cite{koscher2010experimental,miller2013adventures,checkoway2011comprehensive} and then demonstrated feasible from remote~\cite{miller2015remote,nie2017free,Keen2018experimental}, and consist mainly in forging packets from exploited or malicious \acp{ECU}, which limits the capabilities of the attacker to those attacks that can be implemented through sending valid \ac{CAN} frames on the bus. More recent attacks, however, such as the one described in~\cite{palanca2017stealth}, have demonstrated that vulnerabilities also exist at the \ac{CAN} data link layer. These attacks are more powerful and harder to detect, but they are feasible only given that the attacker can inject bits that break the protocol rules at a low level (e.g., if it is possible to write on the bus while another \ac{ECU} is writing at the same time).
In practical scenarios, this requires high precision, especially on a high-speed CAN bus, which is hardly possible on a resource-constrained microcontroller, such as those found in ECUs.
Hence, up to now, link-layer attacks against \ac{CAN} were considered feasible only if the attacker had physical access to the \ac{CAN} bus, with the exception of the work by Kulandaivel et al.~\cite{Kulandaivel2021cannon}, which, however, comes with significant limitations.

In this paper, we take a step forward by presenting \systemns, a novel approach to link-layer attacks against \ac{CAN} that exploits the presence of peripherals connected to the same physical pins of the \ac{CAN} controller (i.e., pin conflicts) to send and receive bits on the \ac{CAN} bus on its behalf. These peripherals can be enabled and disabled by accessing dedicated memory-mapped registers from software, making this approach completely applicable without any hardware modification.
Since conflicting peripherals cannot handle complete \ac{CAN} frames out-of-the-box, we also introduce the concept of polyglot frames, which is inspired by the general notion of polyglots that has been applied in other fields (e.g., polyglot programs, files, signals)~\cite{bratus2015fillory, albertini2014pdf}. This enables us to identify and produce sequences of bits that are compliant with both the \ac{CAN} protocol and the involved peripherals, and generate data sequences that can be transmitted by a protocol interface and can be interpreted as a valid message by a different one.

We demonstrate the validity of our approach from different perspectives. First, we verify the presence of pin conflicts between the embedded CAN controller and other less constrained peripherals, such as UART, SPI, I2C, and ADC peripherals, in a variety of automotive-graded microcontrollers and the existence of corresponding polyglot frames.
Secondly, we demonstrate that the aforementioned peripherals can be practically used to produce and receive complete CAN frames on a real CAN network, keeping up with the speed of the modern CAN bus. This shows the flexibility of our approach and demonstrates practically that real CAN hardware cannot distinguish \system bits from legitimate CAN frames.  

Finally, we implement an end-to-end targeted denial-of-service attack using \systemns, showing how a remotely compromised ECU can completely shut down another ECU in the same CAN network without any assumption on the periodicity of the victim's messages, which is instead a requirement for~\cite{Kulandaivel2021cannon}.

To summarize, our contributions are the following:
\begin{itemize}
\item We present a novel, software-only approach to reliably and precisely read and inject bits on the CAN bus, bypassing the restrictions imposed by the CAN controller;
\item We show that this approach can be used to mount link-layer attacks on CAN networks from remotely compromised ECUs, which makes remote link-layer attacks practical;
\item We demonstrate the possibility of producing full CAN frames that are completely compliant with CAN timing and format specifications, using polyglot frames;
\item Finally, we release\footnote{Available at \url{https://github.com/necst/CANflict}, uploaded for review in the additional materials. \label{url_release}} an extensible framework that can be used to read and write arbitrary bits on the CAN bus using different microcontrollers and peripherals, and can be further extended to include other hardware and protocols for future research on the topic.
\end{itemize}
\section{CAN Protocol Primer}
\label{sec:background}

\ac{CAN} is a bus standard widely used in the automotive industry. 
The ISO 11898 standard~\cite{can-iso2} defines three layers for the CAN protocol stack in relation to the OSI model: the \textit{physical layer}, which defines the electrical properties of the bus, the \textit{data link layer}, which defines frame formats, arbitration, and error reporting mechanisms, and the \textit{application layer}, in which further protocols can define their message formats. Each message at the application layer is composed of a payload, which can be 8 bytes long at most in standard \ac{CAN}, and an ID, which is used as a message identifier and, implicitly, as a priority tag as well, as explained below. 

\mypar{Nodes Layout} At a physical level, \ac{CAN} is a two-wire differential bus that interconnects nodes in a broadcast fashion. Each node that participates in CAN communication requires a CAN interface, which is composed of a \textit{CAN controller} and a \textit{CAN transceiver}.
The \textit{CAN controller} unit can be found as a stand-alone circuit or, more often, as a dedicated module of the host microcontroller. The controller implements the \ac{CAN} protocol at the data link layer as described by the standard, generating the bit sequence that has to be transmitted on the bus and decoding incoming bits into application-level messages. 
The \textit{CAN transceiver} is responsible for converting between logical data, coming out and going to the \ac{CAN} controller, and the corresponding physical signaling, as it connects the \ac{CAN} controller to the physical communication lines.


It is important to note the difference in the signals handled by these two
components: the \ac{CAN} controller is in charge of converting application-level objects (messages) to a sequence of bits and vice-versa, using the \texttt{CANTX} and \texttt{CANRX} lines (which are digital). On the other hand, \ac{CAN} transceivers take care of transforming each bit received from \texttt{CANTX} into a voltage difference between \texttt{CANH} and \texttt{CANL}, and continuously monitor
the bus to output the current differential level as a \texttt{1} or  \texttt{0} on the \texttt{CANRX} line.

\mypar{Bits Representation and Message Arbitration}
The \ac{CAN} specification defines two kinds of bit: \textit{dominant} bits, whose logical value is conventionally  \texttt{0}, and \textit{recessive} bits, corresponding to logical  \texttt{1}. According to the specification, a dominant bus level must always overwrite a recessive bus level. Therefore, the \ac{CAN} bus is implemented as a wired-AND bus. As a result, if any device on the \ac{CAN} bus transmits a dominant bit, which is represented by a \texttt{0} at the logical level, it will overwrite any other ongoing communication. This behavior enables the implementation of \ac{CAN} \textit{arbitration mechanism}. The CAN arbitration mechanism is applied over the first part of the frame, i.e., the ID field. If two or more nodes start sending a frame at the same time, they each continue the transmission as long as the value of the bit read out from the bus equals the value they have written on the bus. Whenever one of the devices reads a value on the bus that is different from the one it has written, it will immediately back off. Since a node putting a recessive bit on the bus will always lose arbitration to a node writing a dominant bit, identifiers with lower values have higher priorities.

\mypar{Error Handling}
The CAN protocol defines an error detection mechanism based on bus monitoring, performed by both the sender and receiver of a message. The sender is responsible for monitoring the sent message bit-by-bit and reading the acknowledge field. Whenever an error is detected, the detecting node starts sending an error frame beginning from the first bit following the error detection. After the error frame is sent and the intermission time has elapsed, the sender of the erroneous message will try to retransmit it. Note that this mechanism happens at the data-link layer: the CAN controller automatically generates error messages and retransmits them after an error has been detected. Therefore, the application layer is never involved in these operations. 

\begin{figure}[tb]
      \begin{center}
            \includegraphics[width=.9\linewidth]{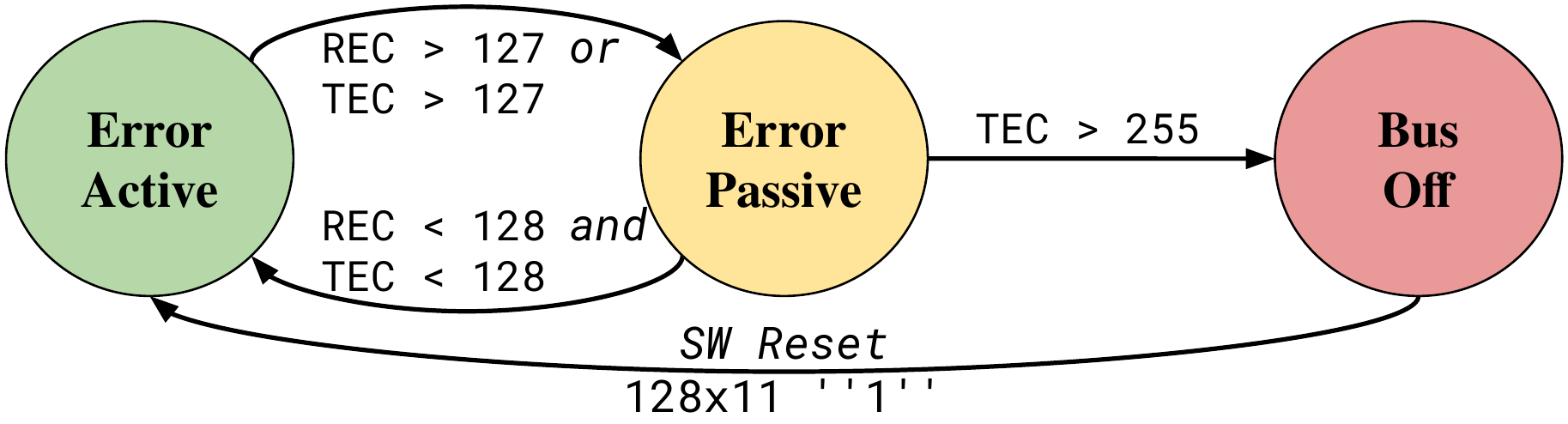}
      \end{center}
      \caption{Error states of the CAN bus.}
      \label{fig:errors-fsm}
\end{figure}

\mypar{Fault Confinement} The CAN protocol specification describes a fault confinement mechanism to prevent faulty nodes from creating high bus loads. According to this mechanism, each node should implement two error counters: \texttt{TEC} (Transmission Error Counter) and \texttt{REC} (Receive Error Counter). These error counters are decremented by \texttt{1} on each successful transmission or reception of a data frame, respectively. Upon detecting an error, the sender node increments \texttt{TEC} by \texttt{8}, while receivers increment \texttt{REC} by \texttt{1} unless they are the ones causing the error, in which case \texttt{REC} is incremented by \texttt{8}. Depending on the values of these error counters, a CAN node can be in one of three error states: 

\noindent \textit{Error Active}: when in this state, the CAN node behaves normally without any specific restriction.

\noindent  \textit{Error Passive}: nodes in this state can only indicate an error by sending 6 recessive bits, preventing other nodes from globalizing the error. When sending consecutive data frames, nodes in this error state must wait for an additional time equivalent to 8 bits (Suspended Transmission Time). 

\noindent \textit{Bus-Off}: nodes that reach the bus-off state can no longer influence the bus communication in any way. This state can only be exited after $128\times11$ correctly recorded recessive bits. 

Figure~\ref{fig:errors-fsm} shows the possible transitions between these three states along with the triggering conditions.

\section{Current State of Attacks on CAN} 
\label{sec:preliminarydiscussion}

Like many older low-level protocols, the CAN protocol lacks some fundamental security mechanisms, making vehicles vulnerable to malicious adversaries. 
As described in \cite{sensors}, \ac{CAN} main security shortcomings are related to the lack of authentication and encryption, the broadcast transmission, the priority-based arbitration, and the limited bandwidth and payload. 
On top of the intrinsic security shortcomings of \ac{CAN}, the environment in which it is most commonly implemented, vehicles, has nowadays multiple attack surfaces, represented by external and internal communication interfaces. In fact, a considerable amount of research has been carried out on this subject~\cite{bozdal2020,bozdal2018,checkoway2011comprehensive,koscher2010experimental,miller2013adventures,miller2014survey}, highlighting many aspects of modern vehicles that can be exploited by malicious actors. 
From a general perspective, it is possible to categorize \ac{CAN} attacks depending on the attacker's location or depending on the network layer at which the attack is carried out. 
Regarding the attacker location, the most common ways to gain access to the \ac{CAN} bus are: \textit{local -} having a malicious node physically installed in the electronic system or attached to the diagnostic port by an adversary, and \textit{remote -} remotely compromising a legitimate internal node of the \ac{CAN} bus.
It is evident that the scale of vulnerable targets in the event of an attack that falls in the second category is much greater, as studied by Miller and Valasek in~\cite{miller2015remote}, where they observed that many hundreds of thousands of vehicles in the united states were vulnerable to their attack at the time of development.
From a network layer perspective, CAN attacks can be carried out at the \textit{application layer} or \textit{data link layer}.

\subsection{Application Layer Attacks}

In standard \ac{CAN} networks, an adversary who is able to attach a malicious \ac{ECU} to the network or reprogram an existing one can typically send and receive messages without any limitations regarding their ID or payload. This capability makes the following attacks possible: 
\textbf{(a)} Eavesdrop Attack - since \ac{CAN} does not implement encryption and has limited payload and bandwidth, it is uncommon for messages to be fully encrypted, which implies that anybody listening on the bus has full read access to the messages sent by all nodes. \textbf{(b)} Spoofing and Replay Attacks - since there is no authentication, an attacker capable of writing a message can impersonate any node, either by forging ID and payload or re-transmitting a previously received message. \textbf{(c)} Network Denial of Service - since \ac{CAN} arbitration depends on the ID, an attacker can overload the bus with \texttt{0x00} ID packets, forcing all other nodes to delay their communication. 

These attacks can be used to target safety-related systems, e.g., the ABS, change the information displayed on the dashboard, either hiding an existing issue to the driver or signaling a nonexistent one, disturb or take control over autonomous features, such as parking assistance or cruise control, and finally completely shut down a car, as famously demonstrated in~\cite{miller2015remote}.
However, since active attacks of this type rely on injecting additional messages on the bus and the vast majority of \ac{CAN} frames are sent with some degree of periodicity~\cite{zago2020recan}, such attacks are trivially detected by modern \acp{IDS}~\cite{al2019intrusion,longari2020cannolo,hanselmann2020canet}.

\subsection{Data Link Layer Attacks}

More recent research has shown that attacks on CAN networks can also be carried out at the data link layer, flying under the radar of message-level \acp{IDS}. Some of the attacks that can be carried out at this level are: \textbf{(a)} Complete Denial of Service - an electrical property of the CAN bus is that dominant bits, i.e., \texttt{0}s, have a priority over recessive bits. This means that keeping the bus constantly in a dominant state will prevent any further communication from being performed on the bus. \textbf{(b)} Selective Arbitration Denial - since messages with lower IDs have a higher priority in the CAN protocol, injecting dominant bits on the bus while a message ID is being communicated will cause the transmitting device to lose the arbitration, which forces it to back off and stop transmitting. This can be done repeatedly during the transmission of specific messages to prevent an ECU from ever winning bus contention. \textbf{(c)} Targeted Denial of Service - if, instead, a dominant bit is injected in the payload of a message, while a transmitting device is sending a recessive bit, the transmitting device will detect an error on the bus, increasing its internal error counter and immediately terminating the transmission. Repeating this process a certain number of times will cause the device to accumulate too many errors, which forces it to go into a bus-off state. This mechanism can be used to completely shut down the communication of any of the nodes connected to the CAN bus. \textbf{(d)} Synchronization Disruption - finally, both the synchronization mechanisms and the sampling point settings of the CAN protocol can be used to cause a desynchronization between nodes on the CAN bus, and, in some cases~\cite{yue2021cancloack, janusattack}, this can cause different nodes to read differently the same message on the bus. This can be particularly useful to evade \ac{IDS} message inspection or disrupt communication between nodes on the \ac{CAN} bus.

While complete DoS attacks are trivial to identify and prevent, more sophisticated attacks such as targeted DoS are extremely hard to distinguish from a genuine fault on the bus and are much more challenging to detect with an \ac{IDS}. However, these attacks come with stronger requirements on the attacker's side. Cho et al.'s approach~\cite{cho2016error} for example, which is based on overlapping a valid message with a forged one to trigger the generation of errors, requires the attacker to be able to predict the arrival of a message with an error of a fraction of the bit time, and recent
work~\cite{kulandaivel2019canvas} demonstrated its unreliability in real-world scenarios. Attacks such as those introduced by Palanca et al.~\cite{palanca2017stealth} and further developed in~\cite{murvay2017dos,bloom2021weepingcan} rely on the ability of the attacker to read the initial part of an incoming message and replace a single recessive bit of the payload with a dominant one, which imposes tight timing constraints, especially at high bitrates. Desynchronization attacks have even stricter timing requirements: in \cite{yue2021cancloack} the attacker has to craft packets in such a way that the rising or falling edge between one bit and the subsequent ones in the CAN frame happen in a time window that is the order of ~1/10 of the bit time (which is already 1$\mu$s on a full-speed CAN bus).
Such hard timing requirements make implementing these attacks on real microcontrollers, such as those found at the core of automotive ECUs, quite challenging and, in some cases, impossible: in \cite{yue2021cancloack}, for instance, the experimental evaluation is carried out using an FPGA rather than a microcontroller, while \cite{palanca2017stealth} uses a 50kbit/s bus as a target for its attack, which is many times slower than the average. A more extensive evaluation of the performance needed for link-layer attacks, such as \cite{murvay2017dos}, shows that the 1$\mu$s bit time
constraint significantly limits the freedom of the attacker.

\subsection{Existing Bit Injection Techniques}
\label{ssec:bit-injection-techniques}

The growing interest in low-level CAN bus manipulation in recent years, especially in the car hacking community, has produced a number of tools that can be used to intercept and produce CAN traffic. In particular, open-source tools such as CANT~\cite{CANT} and CANhack~\cite{CANhack} use bitbanging to access the CAN bus data-link layer. However, CANT is specifically designed for a high-end microcontroller (STM Nucleo-H743ZI2 clocked at 400MHz) and comes with a custom external shield. CANhack, on the other hand, is a MicroPython tool implemented for both the STM32F405 and the Raspberry Pi Pico, but the tool's author explicitly states that the hardware platform in use must be "fast enough to bit-bang CAN," which excludes lower-end microcontrollers. Finally, the tool has been tested only on a 500kbit/s CAN bus.

More recently, CANnon's approach~\cite{Kulandaivel2021cannon}, based on gating the peripheral clock of the \ac{CAN} controller to delay the sending of a dominant bit until the victim frame is transmitted, significantly relaxes the performance requirements of the targeted platform since any microcontroller with an embedded CAN peripheral can inject bits using this technique. However, this approach has a significant limitation since it does not provide a low-level read primitive. As a matter of fact, while the clock is held in the "loading" phase of the attack, the attacker has no feedback on the current state of the bus. This means that the attacker must rely on periodic messages to know when to "fire" the attack and is blind to the traffic that is happening in real-time on the bus. Hence, Targeted DoS attacks mounted with this technique must make heavy assumptions about the current state of the bus before being triggered and rely on specific characteristics of the CAN network traffic, requiring the attacker to predict the time of arrival of a given CAN frame beforehand, which is not always possible nor practical in real-world scenarios.
Finally, this approach is inherently noisy, and the technique discussed by the authors to increase its reliability has the drawback of making it easier to detect for an \ac{IDS} since it requires to hold a dominant state for a prolonged period of time.
\section{\systemns: Polyglot Frames on Conflicting Peripherals}
\label{sec:approach}

Current state-of-the-art tools and techniques used for CAN bit injection are not suitable to make sophisticated data link layer attacks practical on remotely compromised ECUs. They are either too imprecise to handle the timing constraints required by advanced data-link layer attacks, or they require high-end hardware, which might not always be available or exploitable in real in-vehicle networks, or make strong assumptions on specific characteristics of the CAN bus traffic, which are not always verified in practice. In this paper, we present a novel, flexible approach to reliably access the CAN data link layer from software. Our approach can be used even on low-tier microcontrollers without any additional hardware. Moreover, our approach does not make assumptions on the targeted CAN traffic since it provides robust read and write primitives that can be combined to mount current and future attacks to the CAN data link layer from unmodified, remotely compromised ECUs. The added requirement for our approach is that the pins that connect the CAN peripheral to the bus must be accessible by other peripherals on the same microcontroller.
At first glance, this requirement may look limiting, however, this condition is met by the vast majority of modern microcontrollers for the reasons that are explained in Section~\ref{ssec:pinconflicts}. For instance, Table~\ref{tab:micro-conflicts} shows a list of conflicts between some of the most common low-level peripherals and the CAN controller on some popular microcontrollers manufactured by the top vendors in the automotive industry.

\begin{table}[b]
    \caption{A list of conflicts with CAN peripherals found in popular automotive microcontrollers~\cite{digikey-mcus}.}
    \label{tab:micro-conflicts}
\resizebox{\columnwidth}{!}{
    \begin{tabular}{ccccl}
        \cline{1-4}
        \multirow{2}{*}{\textbf{Microcontroller}} & \multirow{2}{*}{\textbf{Vendor}}      & \textbf{\# CAN} & \multirow{2}{*}{\textbf{Conflicts}} & \\ 
        & & \textbf{Devices} & & \\
        \cline{1-4}
        V850ES/JC3-H             & Renesas              & 1                       & UART, I2C, GPIO                        & \\
        MPC5554                  & NXP                  & 3                       & SPI, GPIO                              & \\
        AT90CAN32                & Atmel                & 1                       & Timer, GPIO                            & \\
        SPC564A80B4              & ST  Microelectronics & 3                       & SPI, eSCI, GPIO                        & \\
        C8051F50x                & Silicon Labs         & 1                       & SPI, I2C, LIN, GPIO                    & \\
        AURIX TC399XP            & Infineon             & 4                       & SPI, UART, I2C, ADC, GPIO             & \\
        STM32L562                & ST Microelectronics  & 1                       & SPI, UART, I2C, GPIO                  & \\
        \cline{1-4}
    \end{tabular}
    }

\end{table}

\subsection{Pin Conflicts}
\label{ssec:pinconflicts}

At the heart of automotive \acp{ECU}, similarly to any other embedded system, we find \textit{\acfp{MCU}}. Differently from microprocessors, \acp{MCU} are cheap, small, low-power, and specialized computing devices that come with simpler CPUs, on-chip memory, and many hardware peripherals, all baked in the same silicon die. They are also typically very limited in memory size and performance: a typical \ac{MCU}'s clock frequency can range from as little as tens of kHz to a few hundred MHz on high-end devices. On the contrary, consumer electronics CPUs nowadays are typically clocked at a speed of several GHz.
To cope with the need for real-time responsiveness even with such limited hardware, microcontrollers are equipped with a variety of on-chip \textit{peripherals}, which are designed to efficiently implement some specific, commonly-needed functionality, such as handling SPI or I2C packets, converting analog signals at a high frequency, or reacting to changes in external signals.
The growing need for faster communication and the increased number of complex protocols to handle, along with the desire of vendors to provide general-purpose products, has caused modern \acp{MCU} to be packed with many of such hardware peripherals, which need more hardware pins than there are available on the physical package. Moreover, each peripheral is typically connected to multiple sets of pins to provide maximum flexibility to customers during the PCB design and routing phases. As a consequence, many peripherals end up sharing the same pins, whose physical connections are internally multiplexed and redirected to the chosen peripheral through a set of \textit{memory mapped registers}, which can be read and written by software. We refer to these pins as ``conflicting pins'' and the involved peripherals as ``conflicting peripherals''.

\subsection{Bypassing the CAN Controller}

\begin{figure}
    \begin{center}
        \includegraphics[width=.9\linewidth]{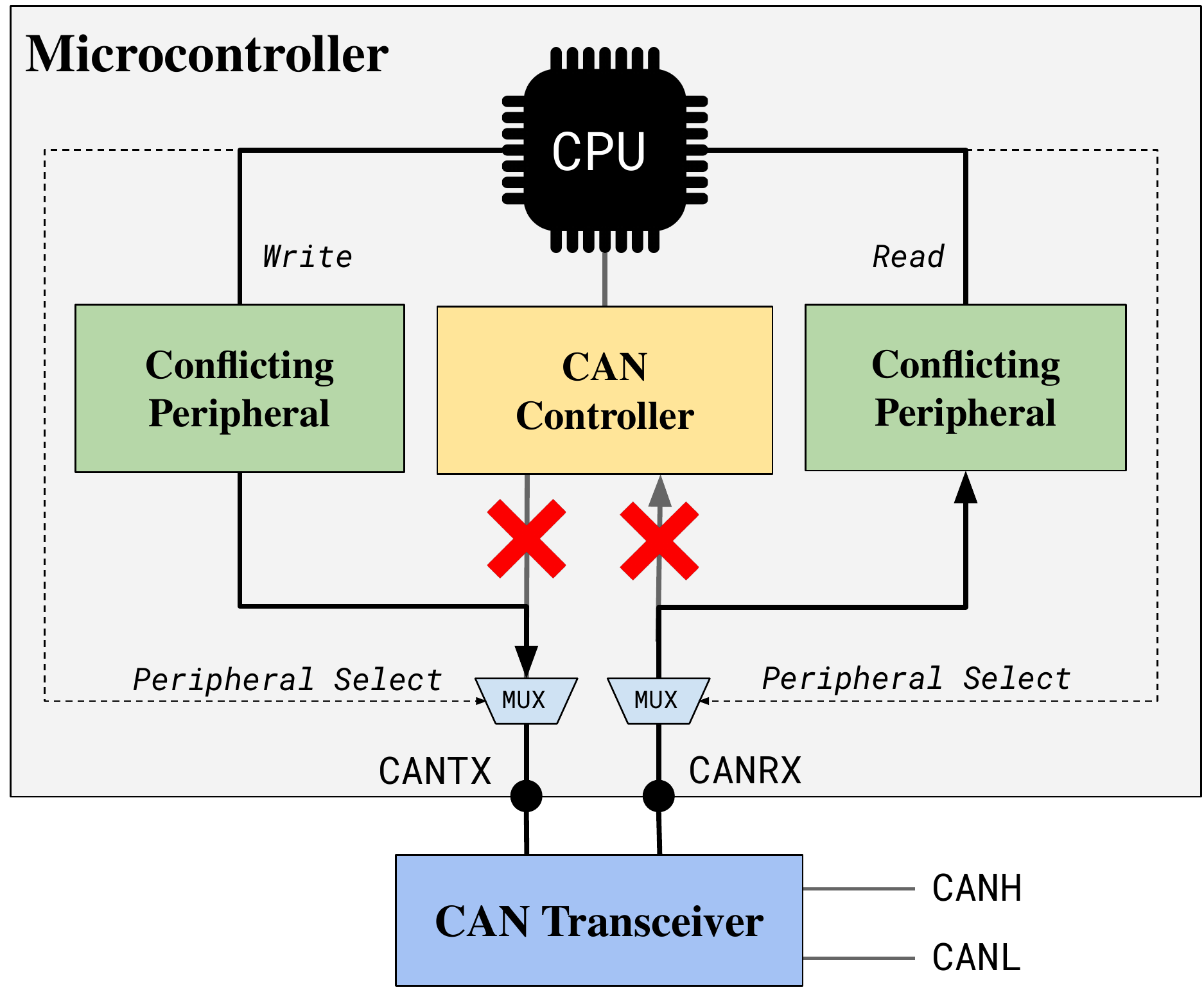}
    \end{center}
    \caption{Conflicting peripheral approach to bypass the \ac{CAN} controller. 
    }
    \label{fig:conflicting-peripherals}
\end{figure}

As explained in Section~\ref{sec:background}, nodes in a \ac{CAN} network access the bus through two components: a controller, which handles the data link layer, and a transceiver, which handles the physical layer.
To perform a link-layer attack, we need to access the CAN data link layer, which, however, is hindered by the \ac{CAN} controller. Indeed, through the \ac{CAN} controller, it is not possible to directly handle each bit that is sent or received on the bus, nor is it possible to force the timing of reading and writing events from software. As a matter of fact, the software layer can only communicate to the \ac{CAN} controller the ID and payload of the message it wants to send or get notified when a complete message is received without errors on the bus, while all the bus arbitration, error handling and frame crafting logic are handled automatically by the hardware. This makes link-layer attacks on \ac{CAN} not feasible through the \ac{CAN} controller. This may create a false sense of security at design time, built upon the assumption that an attacker cannot access the data-link layer of the protocol. However, \ac{CAN} controllers are typically found as embedded peripherals in modern \acp{MCU}, which means that the same pins that are used by the \ac{CAN} controller, \texttt{CANTX} and \texttt{CANRX}, are also shared with other peripherals, as discussed in \ref{ssec:pinconflicts}.

One way we could access these pins is through the \ac{GPIO} peripheral, which enables the software to control and read the logical level of a pin through memory-mapped registers. This technique is commonly called \textit{bitbanging} and can be used on high-end microcontrollers to access a relatively low-speed bus. The main limitation of this technique is that a read or write event for each bit must be commanded by the software, which, in the case of a full-speed CAN bus, leaves the CPU with a 1$\mu$s window to execute whatever logic is needed for each bit, e.g., saving the current level of the bus in a given sequence, counting how many bits have been received and comparing the received sequence with a given one. This is clearly a strong timing limitation for a microcontroller with a clock of, for example, 10MHz. Moreover, with this technique, the CPU is fully occupied by the reading and writing operations, and any event that can alter the timing of an instruction, such as an \ac{IRQ}, might cause the microcontroller to desynchronize with the bus and the attack to fail or be triggered at the wrong moment. Finally, on a practical side, bitbanging techniques rely either on the presence of a high-resolution timer or on platform-specific fine tuning that might not always be practical and is highly prone to problems such as clock drifts.

Instead, \system uses low-level protocol peripherals to control the \texttt{CANRX} and \texttt{CANTX} pins. Our intuition is that we can leverage the presence of conflicting peripherals on those pins to bypass the \ac{CAN} peripheral and gain full access to the \ac{CAN} data-link layer from software, as depicted in Figure~\ref{fig:conflicting-peripherals}. In particular, packets in low-level protocols such as \ac{SPI} and \ac{UART} have much fewer restrictions than \ac{CAN} frames and can be concatenated to overlap partially or completely a given \ac{CAN} message. In this way, we can benefit from the speed and the asynchronous nature of dedicated peripherals to offload the handling of high-speed communications from the CPU, which by itself would not be able to cope with the tight timings imposed by CAN, while still obtaining a high degree of control over the traffic on the bus. The principle described above extends the threat model that car manufacturers should consider while designing vehicular on-board networks by adding this additional attack surface (i.e., remote data-link layer attacks from the application level) that 
is often neglected in existing threat models~\cite{al2019intrusion,longari2020cannolo}.

\subsection{Polyglot Frames}
\label{sec:approach-plyglot2can}

Bypassing the CAN controller is not, by itself, sufficient to mount sophisticated link-layer attacks since, in order to read and write long sequences of bits, we need to handle CAN data from within another peripheral. For this reason, we introduce the concept of \textit{polyglot frames}. In general, the meaning of a signal is not intrinsic to the signal itself (both for digital, as in our case, or analog ones). A digital signal is merely a temporal sequence of high and low voltages, while the actual information stored in the signal depends on the way it is interpreted, following rules and conventions. Therefore, its meaning is not bound to the physical signal itself but to the interpreter that attributes meaning to it. 
For example, a sequence of bytes may output a melody if interpreted as an audio track, but it may look random and meaningless if interpreted as an image. Instead, a sequence of bytes valid both as a sound and an image file is referred to as a \emph{Polyglot file}~\cite{corkami, albertini2014pdf}. 
Following the same reasoning, a bitstream on a communication channel (i.e., frame) transmitted by a protocol interface that is both a valid \ac{SPI} message and a valid \ac{CAN} message would be referred to as a \emph{polyglot frame}.

In practice, we use this concept to identify pieces of CAN frames that are compatible with a given protocol, which can then be read and written with the corresponding peripheral. We also leverage the existence of full-frame polyglots between the protocols chosen in this paper and the CAN protocol to demonstrate that \system can also generate complete and valid CAN frames that are accepted as such by legitimate nodes. 

\subsection{Exploiting Conflicting Peripherals}

Depending on the peripheral that conflicts with the \ac{CAN} controller's \texttt{RX} and \texttt{TX} pins, the capabilities of \system vary. Each protocol has its own requirements, and since it is necessary to fulfill them to read or send a CAN frame, this may limit or fully enable the attacker's capabilities. 
We proceed to analyze the capabilities of our intuition on some of the most common peripherals found on modern microcontrollers, i.e., SPI, UART, I2C, and ADC.

\begin{figure}[tb]
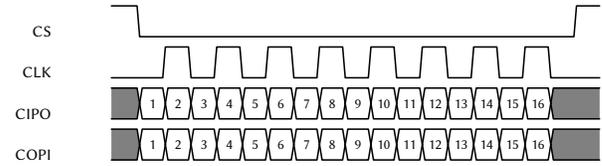


        \newcounter{mycount}
        \tikztimingmetachar{Q}{D{\stepcounter{mycount}\arabic{mycount}}}
        \tikztimingmetachar{R}{[| /utils/exec=\setcounter{mycount}{0} |]}
        \begin{tikztimingtable}[%
                timing/dslope=0.2,
                timing/.style={x=2.5ex,y=3ex},
                x=5ex,
                timing/rowdist=4ex,
                timing/name/.style={font=\sffamily\scriptsize},
            ]
            CS & H 17{L} H \\
            CLK & LL 15{T} LL \\
            \MISOns & U R 8{2Q} 2U \\
            \MOSIns & U R 8{2Q} 2U \\
        \end{tikztimingtable}

    \caption{Timing diagram of an SPI message.}
    \label{fig:spi}
\end{figure}

\mypar{\acf{SPI}} \label{subsec:SPI} SPI is a \master-\slave\footnote{\label{note:masterslave} Although the OSHWA suggests~\cite{oshwa} \emph{Controller} and \emph{Peripheral} as the new naming convention, we avoid the use of this terminology because it could lead to confusion in our context of microcontrollers. Hence, we will refer to the controller device as \emph{\masters} and to the controlled device as \emph{\slave} throughout the text. We will, however, use the suggested acronyms \MISO and \MOSI in lieu of MISO and MOSI for ease of reference.}~serial protocol that typically employs four lines (See Table~\ref{tab:spilines}). Figure~\ref{fig:spi} represents the timing diagram of a typical SPI communication.
When in \masters mode, the SPI device automatically selects the proper \slave, generates the clock signal, sends bits on the \texttt{\MOSI } line, and reads bits on the  \texttt{\MISO} line. Communication from \slaves devices to the \masters device is also initiated by the \master, which decides which device can communicate on the  \texttt{\MISO} line by setting the \textit{CS} line of the corresponding \slaves low and generating the clock signal. The protocol does not define any intrinsic limitation on the shape of the packets sent and received by the \master. 
Finally, SPI devices typically expose some mechanism to modify the \textit{Clock Polarity} and \textit{Clock Phase} of the signal, which determine how the bits are encoded during the communication. In particular, the \textit{Clock Polarity} affects the logic level of the CLK signal when the peripheral is idle, while the \textit{Clock Phase} defines whether the read/write operation for each bit begins on the rising or falling edges of each clock pulse.

\begin{table}[b]
    \centering
    \caption{SPI Lines Description.}
    \label{tab:spilines}

    \begin{tabular}{cp{7cm}}
        \hline
        \textbf{Name} & \textbf{Description} \\ \hline
        \textbf{CS} & Chip select line used by the \masters to select which \slaves to communicate with.\\ \hline
        \textbf{CLK} & Clock signal generated by the \masters and sets the bit timing of the communication. \\ \hline
        \textbf{\MISOns\footref{note:masterslave}} & Data from \slaves to \master. \\ \hline
        \textbf{\MOSIns\footref{note:masterslave}} & Data from \masters to \slave. \\ \hline
    \end{tabular}
\end{table}

\textit{SPI Polyglot Frames}: The main requirement to read on the bus is that the \texttt{\MISO} line of the SPI peripheral and the \texttt{CANRX} line of the CAN peripheral share the same pin so that the incoming signals from the CAN bus can be redirected to the SPI peripheral. An additional requirement is the knowledge of the baudrate of the \ac{CAN} bus. Given these two requirements, it is trivial to read bits on the bus. Typically, to read a complete CAN message from the start, the application code needs to recognize the \acf{SoF} bit. 
Similarly, to write on the bus from the SPI peripheral, the  \texttt{\MOSI } line of the SPI peripheral and the \texttt{CANTX} line of the CAN peripheral need to share the same pin, alongside the knowledge of the baudrate. 
Since no specific rules apply when transmitting SPI packets from a \masters device, any bitstream that is provided to the peripheral will be transmitted as-is on the  \texttt{\MOSI } line. This means that if the  \texttt{\MOSI } line conflicts with the \texttt{CANTX} signal coming out from the microcontroller, the SPI peripheral can be used to send an arbitrary number of bits of the bus. 
Note that, since the \texttt{CS} line and the \texttt{CLK} signal are generated automatically by the SPI peripheral, we can completely ignore their presence. In fact, neither of these signals are relevant for sending arbitrary bits on the CAN bus, and the SPI device never reads them. 
Figure~\ref{fig:allpolyglots} shows an example of how a sequence of bits transmitted by the SPI peripheral can be interpreted both as a CAN message and an SPI message.

\begin{figure}
    \begin{center}
        \includegraphics[width=1\linewidth]{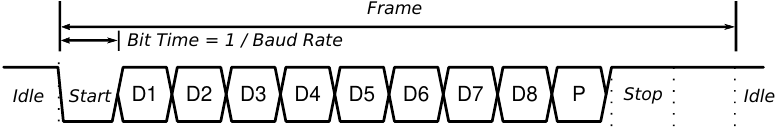}
    \end{center}
    \caption{Timing diagram of a UART message.}
    \label{fig:uart}
\end{figure}

\mypar{\acf{UART}} \label{subsec:UART} The UART protocol is another widespread serial protocol used in many embedded applications. Unlike SPI, the UART protocol does not use a clock signal to synchronize the transmitter and receiver devices; instead, it transmits data asynchronously. Similar to the CAN peripheral signals, the two main signals of a UART peripheral are the transmission line (\textit{TX}) and receiving line (\textit{RX}). In the UART protocol, each packet must have a predefined form, which consists of a start bit, data frame, a parity bit, and stop bits, as summarized in Table~\ref{tab:uartlines} and shown in Figure~\ref{fig:uart}. In particular, the start bit's value is always \texttt{0}, the stop bit's value is \texttt{1}, and the payload of each packet can contain 5 to 9 bits.

\begin{table}[b]
    \centering
    \caption{UART frame Description.}
    \label{tab:uartlines}
    \begin{tabular}{cp{6cm}}
        \hline
        \textbf{Name} & \textbf{Description} \\ \hline
        \textbf{Start Bit} & Always set to \texttt{0}. \\ \hline
        \textbf{Data Frame} & Payload can be from 5 to 9 bits long. \\ \hline
        \textbf{Parity Bit} & Optional, used for error detection. \\ \hline
        \textbf{Stop Bit(s)} & One or two consecutive logical \texttt{1}s, depending on peripheral configuration.\\ \hline
    \end{tabular}
\end{table}

\textit{UART Polyglot Frames:} If there is a pin conflict between the \texttt{CANTX} line and the \ac{UART} peripheral's \texttt{TX} line, we can inject bits on the \ac{CAN} bus using the \ac{UART} peripheral. Equivalently, a pin conflict between the \texttt{CANRX} line and the \ac{UART} \texttt{RX} line can be exploited to read bits on the bus. 
Although the \ac{UART} peripheral still enables an attacker to send and receive bits on the bus at arbitrary moments, bypassing the arbitration logic, the fact that \ac{UART} packets have fixed start and stop bits imposes some additional constraints with respect to SPI when trying to emulate \ac{CAN} traffic. In particular, while the payload of each frame is entirely controlled by software, start and stop bits in each packet have fixed values. However, by modifying the packet length, the user can still control the position of these fixed values, which significantly relaxes the practical constraints of emulating and intercepting CAN traffic. In practice, as shown in Section~\ref{sec:experiments}, by using a chain of \ac{UART} packets, it is still possible to match a significant portion of a CAN message, and, in some cases, we can even craft complete CAN frames by just concatenating a sequence of \ac{UART} packets. Figure~\ref{fig:allpolyglots} shows a particular instance of a \ac{UART} polyglot frame, in which the same signal can be interpreted as both a \ac{CAN} frame and a sequence of \ac{UART} packets. We leverage this property in our experimental validation, proving that a real UART peripheral can generate a signal that is accepted and acknowledged by an unmodified \ac{CAN} controller. 

Given a \ac{CAN} frame that we want to produce, to obtain a valid \ac{UART} polyglot, we follow a ``greedy'' approach: \textbf{(a)} Assign the first packet's length, verifying that the last bit is compliant with the stop bit value. \textbf{(b)} If this is not the case, modify the length until the first packet is a valid \ac{UART} packet. \textbf{(c)} Verify that the following bit is compliant with the start bit value. If this is not the case, repeat from \textit{b.} until this condition is met. \textbf{(d)} Repeat from \textit{a.} for the next UART packet, until all the CAN frame has been covered. \textbf{(e)} If no correct solution can be found, backtrack to reassign earlier packets to a different length. Additionally, the stop bit length of each packet can be modified to provide even more solutions.

For what concerns reading, it should be noted that the presence of a start bit in the \ac{UART} protocol, which is needed to synchronize the communication in the absence of a separate clock line, is actually an advantage for our purpose since it has the same value of the \ac{CAN} frame \ac{SoF} bit. This means that the \ac{UART} peripheral can be configured to start reading the bus at any moment of the inter-frame time window, and it will automatically recognize the start of the next \ac{CAN} frame. As we already noted, subsequent bits must be compliant with the \ac{UART} protocol: the next 5 to 9 bits, depending on how the peripheral is configured, are read normally by the peripheral. The following 1 or 2 bits, depending on Stop Bit(s) configuration, must be  \texttt{1}. If not, the UART peripheral discards the incoming message as faulty. The next bit must be a  \texttt{0} (Start Bit), which forces the peripheral to listen for a new packet, and so on. This implies that, in order to configure the \ac{UART} peripheral correctly, the attacker must know in advance (at least part of) the sequence of bits that it wants to target.

\begin{figure}
    \begin{center}
        \includegraphics[width=\linewidth]{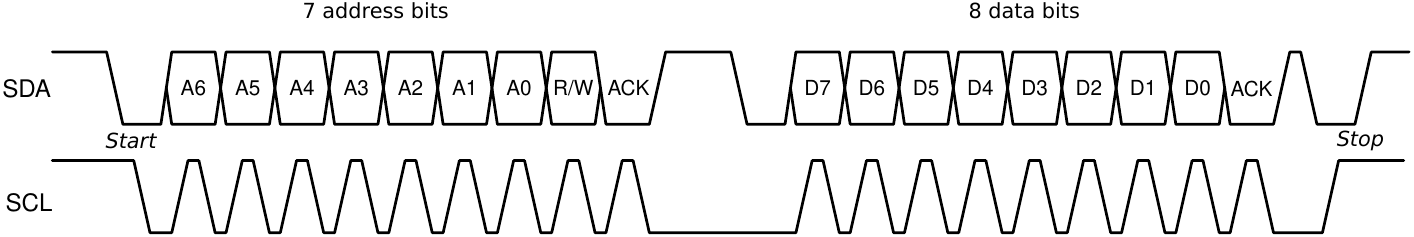}
    \end{center}
    \caption{Timing diagram of an I2C message.}
    \label{fig:i2c-timing}
\end{figure}

\mypar{\ac{I2C}} \label{subsec:I2C}
Another popular communication protocol in modern embedded systems is I2C, which is a multi-\master, multi-\slave\footref{note:masterslave}, synchronous protocol. Since it is a serial protocol, data is transferred bit by bit along a single wire, called the \texttt{SDA} line. Like \ac{SPI}, \ac{I2C} is synchronous. Hence, the emission and sampling of bits are
synchronized by a clock signal shared between the \masters and the \slave. The \masters controls the clock signal.
In the I2C protocol, messages are broken up into two types of frames: an address frame, where the \masters indicates the peripheral to which the
the message is being sent, and one or more data frames, which are 8-bit data messages passed from \masters to peripheral or vice versa.
The two lines are called \textit{Serial Data} (SDA) and \textit{Serial Clock}
(SCL). Data is placed on the SDA line after SCL goes low and is sampled after the SCL line goes high. The time between clock edge and data read/write is defined by the devices on the bus and varies from chip to chip. The timing diagram of a typical \ac{I2C} communication is provided in Figure~\ref{fig:i2c-timing}. \ac{I2C} packets are structured as in Table~\ref{tab:i2c}.

\begin{table}[b]
    \centering
    \caption{I2C communication process.}
    \label{tab:i2c}
    \resizebox{\columnwidth}{!}{
    \begin{tabular}{cp{6cm}}
        \hline
        \textbf{Name} & \textbf{Description} \\ \hline
        \textbf{Start Condition} & The SDA line is pulled low while SCL is high to indicate the beginning of communication. \\ \hline
        \textbf{Payload} & 8 controllable bits on the SDA line, both for address and data frames.\\ \hline
        \textbf{ACK Slot} & The SDA line is held high by the \master , and the \slaves is expected to pull low (\texttt{0}) the clock for a positive acknowledgment. \\ \hline
        \textbf{Stop Condition} & The SDA line is pulled high while SCL is high to indicate the end of communication.\\ \hline
    \end{tabular}
    }
\end{table}

\textit{I2C Polyglot Frames}: If a conflict between the \textit{SDA} line and the \texttt{CANTX} line is
present in the target microcontroller, this protocol can be used to send
bits on the CAN bus. The same does not hold for reading since, in this protocol, the primary must always send a command on the bus before reading.
The protocol rules are even more restrictive than UART since the beginning and end of each packet are signaled by a low voltage on the \textit{SDA} line, whose duration depends on the specific device. Moreover, each packet sent by the \masters is expected to be acknowledged by the \slaves in the \textit{ACK slot}, during which the \masters leaves the \textit{SDA} line in a high voltage state.
Finally, each frame in the I2C protocol is formed by multiple packets and starts with an \textit{address} packet, in which the \masters communicates the address of the \slave, the direction of communication (read or write), and waits for the ACK signal from the \slave. If the message is not acknowledged, the communication is interrupted by the \master. All of these aspects of the protocol interfere with the necessity of sending arbitrary bits on the bus since the voltage of the bus during the start condition, stop condition, ack slot, and inter-frame space cannot be controlled.
Nevertheless, as for UART packets, being able to control at least some of the bits that are sent on the bus can be enough to inject small sequences of bits, and, with enough knowledge of the I2C device characteristics, it is even possible to craft complete, valid CAN messages, as shown in Figure~\ref{fig:allpolyglots}. 

Similarly to UART, the following steps produce a frame valid both for \ac{CAN} and I2C:
    \textbf{(a)} Obtain the durations of the I2C start condition, stop condition, ACK slot, and inter-frame space, either from the datasheet or by direct measurement: these are the fixed portions of the I2C frame.
    \textbf{(b)} Verify that the target \ac{CAN} bit time divides the duration of all fixed portions. This induces a bit representation for each fixed portion.
    \textbf{(c)} Define the number of I2C frames to send. 
    \textbf{(d)} Assign the bits corresponding to the I2C fixed portions, as determined in \textit{b.}.
    \textbf{(e)} Assign the bits that have a fixed value in the CAN protocol (e.g., ACK delimiter).
    \textbf{(f)} Choose the remaining bits of the \ac{CAN} frame, excluding the CRC portion. 
    \textbf{(g)} Calculate the CRC of the frame.
    \textbf{(h)} Check if the CRC is compatible with the already assigned bits of that section of the frame. 
    \textbf{(i)} Repeat from \textit{f.} if the CRC is not compatible.

Note that both \textit{b.} and \textit{e.} may be unfeasible.
On the one hand, given a CAN baudrate there is no guarantee that the I2C writing technique is suited for crafting messages at that baudrate. On the other hand, given a compatible baudrate, the position and length of the I2C fixed portions could interfere with the CAN control fields, leading to a wrongly formatted message. Nevertheless, since many link-layer attacks do not require the ability to craft complete packets, I2C can still be used to attack the CAN data link layer, as demonstrated in Section~\ref{subsec:PoC}.

\mypar{\ac{ADC}}
\label{subsec:ADC}
\textit{\ac{ADC}}s are the last type of peripherals examined in this section. The capabilities of such peripherals may strongly vary among different platforms and vendors, but the basic idea is that they can be used to perform fast and repeated analog-to-digital
conversions without the intervention of the CPU ( e.g., to sample the \textit{CANRX} signal as if it was an analog signal). 
Many microcontrollers include one or more ADC devices as on-chip peripherals. Typically, such devices expose a mechanism to regulate the \textit{resolution} of the conversion, i.e., how small can the difference between two analog values be before they become indistinguishable. Clearly, since we are interested in the digital value of the bus, we can select the lowest possible resolution for these conversions, which typically also means higher sampling frequency, and then compare the result with a constant value, corresponding to half of the full-scale range of the ADC. Analog-to-Digital Converters (ADCs) generally expose a simple interface for sampling analog signals at precise intervals.

\textit{ADC Polyglot Frames:}
Clearly, the only action that can be carried out through an ADC peripheral
is reading. Therefore, if the \texttt{CANRX} signal conflicts with an analog input of the chosen target, the ADC can be used for sniffing bits on the bus. The implementation is straightforward and, in our case, implies the calling of a function every time a new ADC value is received, which checks whether the converted value is greater or lower than the mid-range value of the ADC. We can either listen on the bus until a given sequence of bits is received or record the bus activity for a fixed number of conversions and store the results in a buffer.
\section{Experimental Validation}
\label{sec:experiments}

To validate our approach, we prove that conflicting peripherals can be exploited on real hardware and that they can be used to achieve reliable control of the CAN link layer. In particular, we first demonstrate the practicality of producing CAN polyglots from conflicting peripherals and verify that such polyglots are indistinguishable from legitimate CAN bits by testing the exchange of entire CAN frames between a conflicting peripheral and a real CAN controller and showing that no errors are produced.
We also benchmark the maximum speed at which such full-frame polyglots can be produced on high-, mid-, and low-end microcontrollers, proving that \system provides reliable, high-speed read and write primitives also on low-end microcontrollers.
Additionally, we measure the compatibility of messages coming from real CAN traffic with the constraints described in Section~\ref{sec:approach} for both the UART and the I2C peripheral.
Finally, we demonstrate the possibility of mounting advanced link-layer attacks from a real, unmodified microcontroller with \system by implementing a targeted DoS attack on simulated CAN traffic, which was recorded from a real car. 

\subsection{Experiments Setup}
\label{sec:platforms}

The microcontrollers chosen for our experiments are the NXP LPC11C24~\cite{lpc}, a low-end microcontroller equipped with an ARM Cortex M0 processor, the STM32L562~\cite{stm32l562}, a mid-range controller based on an ARM Cortex M33 processor running up to 110 MHz\footnote{\label{110mhz}In the case of the STM32, since the peripheral frequency is obtained by dividing the system clock by a power of 2 (peripheral prescaler), to get a baudrate compatible with typical \ac{CAN} speeds (e.g., 1 Mbit/s) it is necessary to decrease the system clock speed from 110 MHz to 64 MHz, which can be divided by a power of 2.},  and the Infineon AURIX TC399XP~\cite{tc399}, a high-end, automotive-grade microcontroller with a 6-core processor from the TriCore family. A comparison between the three microcontrollers is provided in Table~\ref{tab:platforms}. The choice of these platforms has a twofold aim. First, we want to demonstrate that the techniques presented in this work are flexible with respect to the specific hardware implementation, making them viable on a huge variety of systems. 
Secondly, we aim at comparing the capabilities of high-, mid-, and low-end platforms, such as the STM32L562, which is found on mid-range systems, and small and inexpensive microcontrollers, such as the LPC11C24, which can be found on simpler systems. 
Since they differ in many aspects, including the vendor, CPU architecture, clock speed, peripheral chips, and overall performance, they are a perfect fit for demonstrating the flexibility of our approach.

We aim to evaluate all peripherals' capabilities on all the platforms considered to produce a fair performance evaluation. However, as shown in Table~\ref{tab:micro-conflicts}, the number of conflicts on CAN peripherals is limited.
To solve this issue, we simulate some of the conflicts by wiring together the signals coming out from the peripheral under test to the \texttt{CANRX} and \texttt{CANTX} signals, which are connected to the CAN peripheral. In other words, we are simulating with external wiring the behavior that is normally displayed by internal signal multiplexing.
Even if this is not the hardware setup that we have considered when devising the conflicting peripheral techniques, we are confident that our setup closely mimics a situation in which two peripherals have a pin conflict in the chip without losing the capability of evaluating their relative performances. A complete demonstration of the feasibility of such techniques on real pin conflicts is provided in the second experiment in Section~\ref{subsec:PoC}.

\begin{table}[b]
    \caption{Comparison between the LPC11C24, STM32L562, and the TC399XP microcontrollers.}
    \label{tab:platforms}
   \resizebox{\columnwidth}{!}{%
        \begin{tabular}{lccc}
            \cline{1-4}
                             & \textbf{LPC11C24} & \textbf{STM32L562} &\textbf{TC399XP} \\ \cline{1-4}
            Vendor           & NXP               & ST Microelectronics & Infineon          \\
            Architecture & Cortex M0 & Cortex M33 & Tricore 32-bit \\ 
            Cores            & 1                 & 1 & 6                 \\
            SRAM Size        & 8 kB              & 256 kB & 2.9 MB            \\
            Flash Size       & 32 kB             & 512 kB & 16 MB             \\
            CAN Peripherals  & 3          & 1 & 3          \\
            SPI Peripherals  & 2                 & 3 & 6                 \\
            I2C Peripherals  & 1                 & 4 & 2                 \\
            UART Peripherals & 1                 & 3 & 12                \\
            ADC Channels     & 8                 & 2 & 12                \\
            CPU Speed        & 50 MHz            & 110 MHz\footref{110mhz} & 300  MHz          \\ \hline
        \end{tabular}
}
\end{table}

\subsection{Full Frame Experiments}
\label{sec:experiments-messages}

To evaluate the reliability of our approach on a high-speed CAN bus, we decided to leverage the existence of polyglot frames for all the chosen peripherals. For each platform, we simulate the exchange of complete CAN frames between an attacker node, which uses \system, and a victim node, which uses a regular CAN controller.
In this way, on the one hand, we can generate an entire CAN frame from a peripheral on the attacker node and verify that the victim node's CAN peripheral accepts it without errors by observing the acknowledgment bit on the bus.
On the other hand, we can test that entire CAN frames are correctly read by the chosen peripherals on a high-speed CAN bus by verifying that the received bits correspond to the CAN frame sent by the legitimate node.
With this strategy, we ensure that all errors and drifts that might accumulate during the transmission or receiving are still smaller than what can be sensed by an ordinary CAN controller.
We believe that, in this way, we can convince the reader of the advantage of using conflicting peripherals over other, less reliable techniques.
We also imply that if we can repeatedly produce and receive full frames at high speed without errors, we can do so also for smaller portions of the CAN message, which is the typical requirement for data link layer attacks.

\begin{figure}
    \centering
    \resizebox{!}{0.9\height}{
    \includegraphics[width=\linewidth]{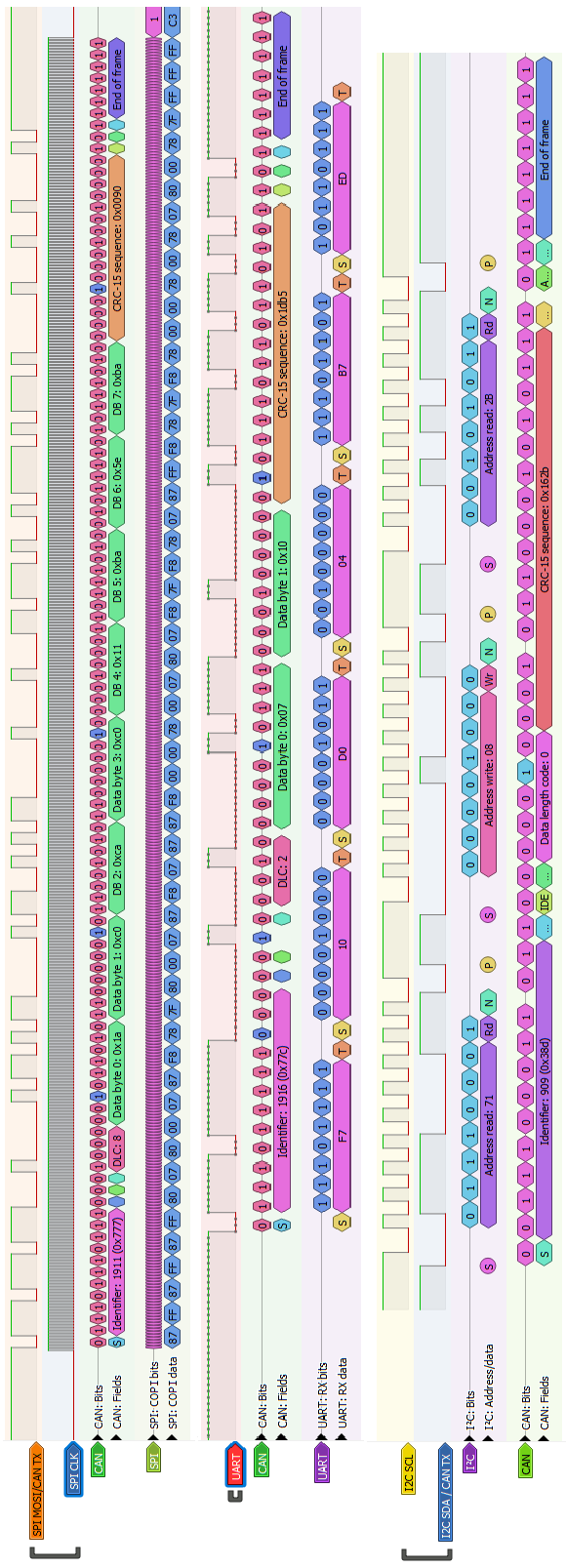}
    }
    \caption{Three examples of how SPI , UART, and I2C peripherals can be employed to craft valid CAN messages.}
    \label{fig:allpolyglots}
    \end{figure}

More specifically, in our experiments, we used three different frames, a generic one and two specifically crafted ones for the UART and I2C peripherals. The bit representation for each frame can be found in Figure~\ref{fig:allpolyglots}.

\mypara{Generic Frame} The generic frame sent and received through the SPI interface is a CAN frame with the longest possible payload (8 bytes) and generic content. This frame has also been used to validate the ADC reading. It has no particular features and represents a randomly picked, standard CAN frame. 

\mypara{UART Frame} The CAN frame chosen to test the UART peripheral has been instead crafted to respect both the restrictions of CAN frames and those of the UART protocol, as discussed in Section~\ref{subsec:UART}. In particular, the chosen frame has a 2-byte payload, and the ID and content of the payload have been chosen in such a way that all the fields, including the CRC field, do not break the UART rules.

\mypara{I2C Frame} The CAN frame sent by the I2C peripheral is a simple remote frame request with a null payload and an ID of \texttt{0x38d}. Similar to the previous frame, its values have been selected to be compliant with the strict rules imposed by the I2C protocol, described in Section~\ref{subsec:I2C}.
Figure~\ref{fig:allpolyglots} contains the bit representation of this frame, as well as the interpretation for both the CAN protocol and the I2C protocol. In this figure, each packet's payload is colored in violet, while the start condition duration is represented by the space between the pink circle (\texttt{S}) and the start of the payload. Other fixed bits are the ACK bit (green segment), which has to always be \texttt{1} since no I2C \slave~is connected to send an acknowledgment, and the stop condition, which is the space between the end of the \texttt{ACK} bit and the yellow circle (\texttt{P}). The inter-frame space, which is the time between the stop condition of a packet and the start condition of the following one, is also fixed and must be \texttt{1}.

As a baseline for our experiments, we use a custom implementation of the bitbanging technique, specifically tailored for each given platform and message. This has been produced using both high-resolution hardware timers and busy-wait loops on the CPU and optimized manually with many trial-and-error attempts in order to show the maximum performance that can be extracted by the platform's CPU for this task. Since such fine-tuning requires physically measuring the accumulated drift for every bit on the bus with an oscilloscope and manually compensating it in the code, we consider this useful only for benchmarking purposes. In fact, an attacker in a remote scenario would not have access to such detailed information about the bit timings on the bus.

\subsection{Results}

Table~\ref{tab:results} reports the maximum bus speed at which we were able to send or receive 100 consecutive CAN frames without any errors on the bus. We observe that SPI and UART peripherals are particularly fit for achieving reliable, high-speed communication on the CAN bus, even on low-end microcontrollers such as the LPC11. I2C peripherals, on the other hand, are less useful in full-frame generation and reception due to the restrictions imposed by the protocol, but they can be nevertheless used to generate shorter sequences, as shown in Section~\ref{subsec:PoC}. 
Since the AURIX TC399XP and STM32 L562 MCUs are significantly more powerful than the LPC microcontroller, we were able to achieve the highest possible bitrate also with our custom bitbanging implementation. However, even on such platforms, reading bits with bitbanging and busy-wait loops was proven to be less reliable than writing since some packets were incorrectly read at a baudrate of 1 Mbit/s. 
The main reason is that, while during writing, the transmitting device imposes the timing of the communication to other devices using the \textit{soft resynchronization} mechanism, during reading operations, the device does not have this power. In our case, since the interval generated by software between one bit and the other was slightly more than 1 $\mu$s, this difference was accumulated during the sampling of the packet until a bit was incorrectly read. Decreasing the interval between bits caused instead the bit timing to be much lower than 1 $\mu$s and bits to be read twice. Using the high-resolution hardware timer to resynchronize the communication periodically was enough to cope with these small timing deviations on the TC399. 

\begin{table}[b]
    \centering
    \caption{Performances of peripherals w.r.t each boards of our experiments in write (W) mode or read (R) mode. \texttt{-} refers to cases with impossible setups, while \texttt{n.a.} refers to implementations that are theoretically feasible but our framework does not support yet.}
    \label{tab:results}
\resizebox{\linewidth}{!}{
    \begin{tabular}{c|cccccc}
        \hline
        \multirow{2}{*}{\textbf{Platform}} & \multicolumn{2}{c}{\textbf{LPC11C24}} & \multicolumn{2}{c}{\textbf{STM32L562}} & \multicolumn{2}{c}{\textbf{TC399XP}} \\
        & W & R & W & R & W & R \\
        \hline
        \textbf{Bitbanging} & 200 kb/s & 120 kb/s & 1 Mb/s & 500 kb/s & 1 Mb/s & 1 Mb/s  \\
        \textbf{SPI} & 1 Mb/s & 1 Mb/s & 1 Mb/s & 1 Mb/s & 1 Mb/s & 1 Mb/s \\
        \textbf{UART} & 1 Mb/s & 1 Mb/s & 1 Mb/s & 1 Mb/s & 1 Mb/s & 1 Mb/s  \\
        \textbf{I2C} & 200 kb/s & - & 100 kb/s &  - & n.a. & - \\
        \textbf{ADC} & - & \textless50 kb/s & - & 300 kb/s & - & 1 Mb/s \\
        \hline
    \end{tabular}
}
\end{table}

On the other hand, in the LPC microcontroller, the difference between bitbanging and our approach is more evident, as the CPU alone cannot cope with the speed of the bus. This is a significant result since it shows how peripheral-based techniques enable previously impossible precision in injecting and reading bits on the bus. In particular, on this platform, writing techniques that employed hardware peripherals performed 5 to 10 times better than the basic bitbanging implementation, reaching the maximum bus bitrate of 1 Mbit/s, while the bitbanging techniques could not be faster than 200 kbit/s. This demonstrates the increased reliability of our techniques with respect to traditional bitbanging on such platforms.

Finally, due to the restrictions that the protocol imposes, the I2C technique was implemented only for a specific bitrate (200 kbit/s for the LPC and 100 kbit/s for the STM32 platforms). To better explain I2C restrictions, the measured timings for the fixed portions of the frames for the LPC platform were: 5.2 $\mu$s (start), 4.41 $\mu$s (ack), 5.33 $\mu$s (stop), and 9.58 $\mu$s (interframe space), with an expected error of $\pm$0.25 $\mu$s. Therefore, a bit time of 5 $\mu$s (equivalent to a \ac{CAN} baudrate of 200 kHz) can contain start, ack, and stop conditions while the interframe space is contained in 2 \ac{CAN} bits. Empirically, we discovered that at this baudrate (200 kHz), the deviations of I2C from the nominal bit time are acceptable, and the error w.r.t. the nominal bit time is absorbed by the synchronization mechanisms of \ac{CAN}. This does not hold for multiples of this baudrate since the \ac{CAN} bit time becomes smaller (e.g., 2.5 $\mu$s for a baudrate of 400 kHz), and therefore, the relative error generated by the deviations of the I2C fixed portions increases.

\subsection{Targeted Denial of Service Experiments}
\label{subsec:PoC}

To demonstrate the capabilities of \system in real-world scenarios and its adaptability in terms of possible peripheral combinations, we implement a targeted denial of service attack against a busy \ac{CAN} network with a baudrate of 1Mb/s, using an \ac{SPI} peripheral to read and an \ac{I2C} peripheral to write on the bus.

\begin{figure}[tb]
    \centering
    \includegraphics[width=\columnwidth]{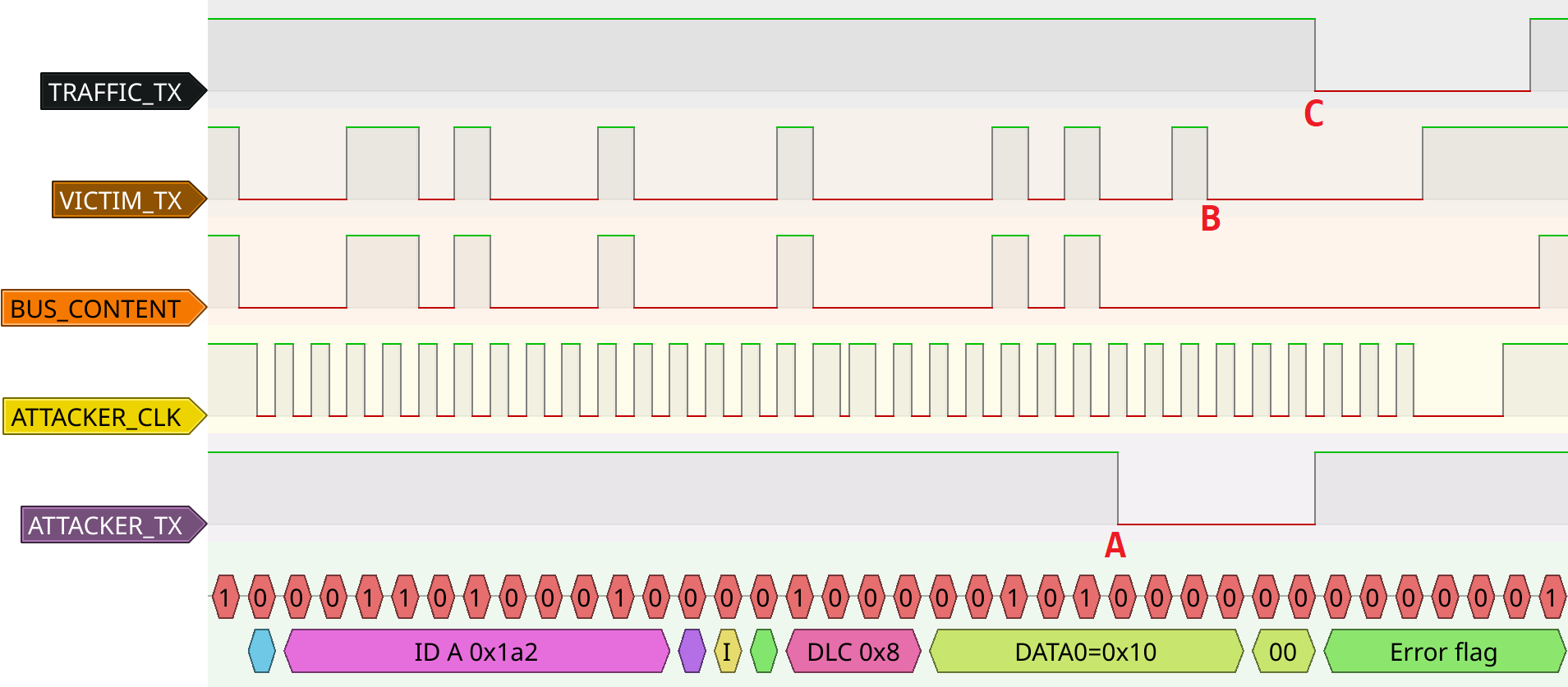}
    \caption{Trace representing an attacker triggering an error on a victim's frame. The attacker is using SPI to read and I2C to write on the bus. The attack starts at A, triggers an error flag by the victim at B, and by the rest of the bus at C.}
    \label{fig:targetedattack}
\end{figure}

We choose the STM32 L562 as the attacker, the AURIX TC399 first CAN line as the victim, and its second CAN line as a traffic generator (the AURIX board has two physically separated, completely independent \ac{CAN} nodes). The traffic chosen to simulate a real vehicle was retrieved by the ReCAN dataset~\cite{zago2020recan}. One of the IDs (\texttt{0x1A2}) is chosen to be the victim frame and is transmitted by the victim node, while the rest of the dataset is being transmitted by the traffic generator node. 
The attacker uses the \ac{SPI} peripheral, connected to the \texttt{CANRX} pin, to scan the bus searching for the victim's ID, and the \ac{I2C} peripheral, connected to the \texttt{CANTX} pin, to write a sequence of dominant bits to trigger the detection of an error by the victim and its fault confinement mechanism. In Figure~\ref{fig:targetedattack}, we see a trace captured while the attacker triggers an error on the bus while the victim is communicating: The attacker forces the bus in the dominant state for the duration of 6 bits (A) while the victim sends its payload, triggering an error flag from the victim (B), and from the rest of the nodes on the bus at (C). 

The results of the experiment are optimal: at the highest \ac{CAN} baudrate of 1Mb/s, the STM32 has no issues detecting the ID of the victim through the \ac{SPI} peripheral and then generate the error through \ac{I2C} one. This process is executed 32 times to trigger the fault confinement mechanism of the victim and prevent it from communicating on the bus. The attacker never erroneously generates an error on a frame with another ID. The experiment was repeated multiple times to ensure consistency of the results. 

\subsection{Polyglot frames compatibility in CAN traffic}

\begin{figure}
    \centering
    \begin{subfigure}{\linewidth}
    \centering
    \includegraphics[width=.9\columnwidth]{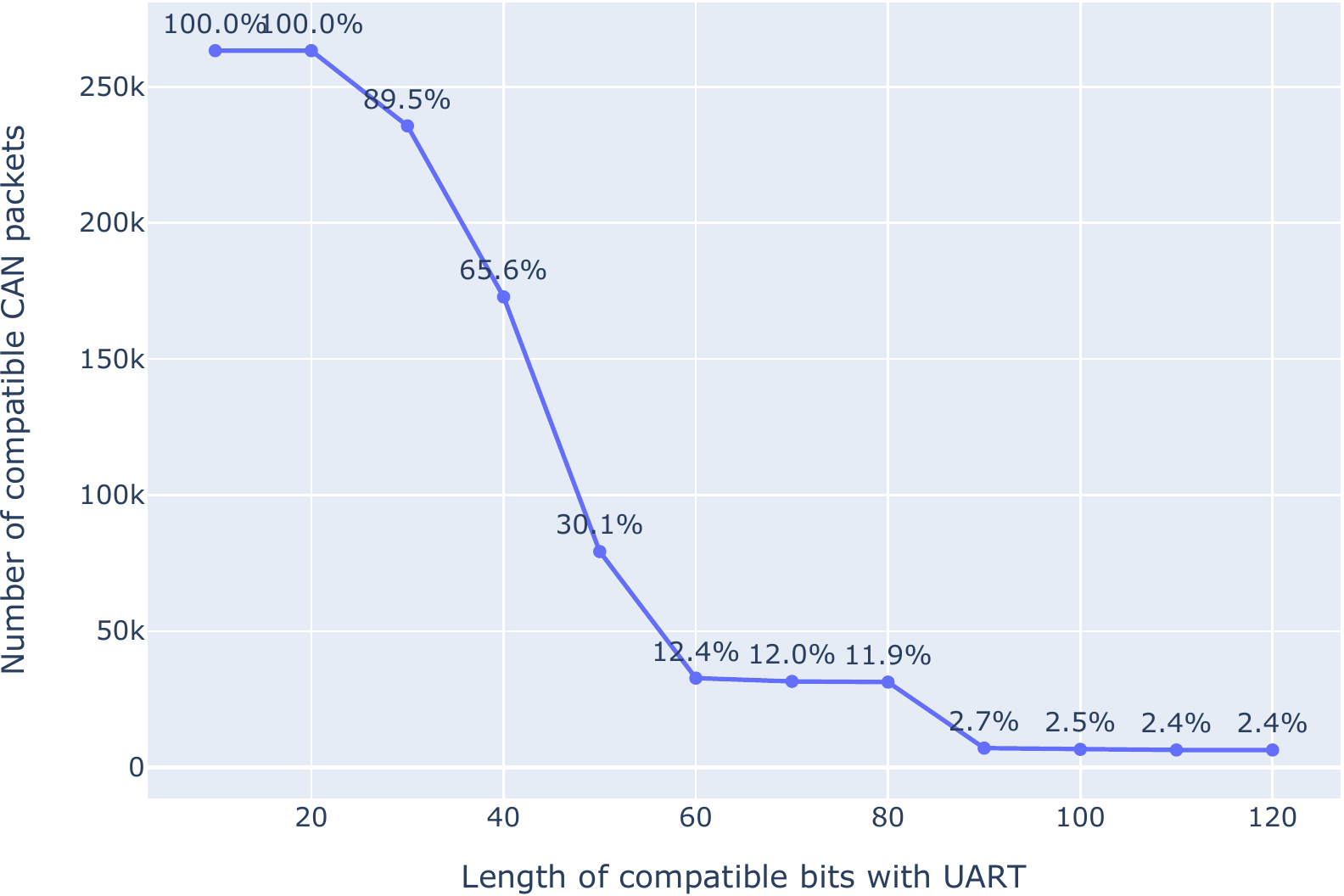}
    \label{fig:uartcompatibility}    
    \end{subfigure}
    \begin{subfigure}{\linewidth}
    \centering
    \includegraphics[width=.9\columnwidth]{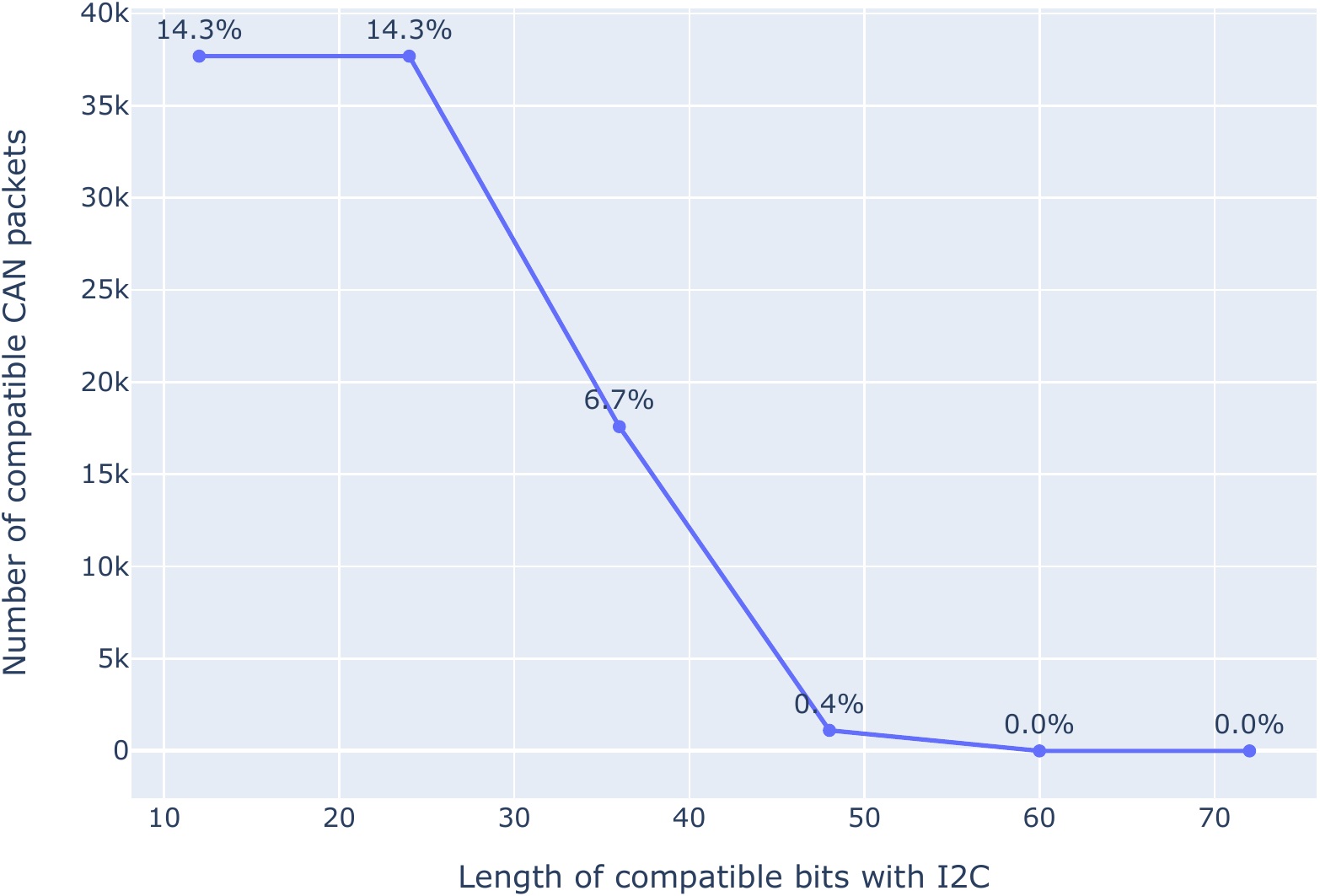}
    \label{fig:i2ccompatibility}    
    \end{subfigure}
    \caption{Number of CAN messages taken from a dataset of ReCAN\cite{zago2020recan} are compatible polyglots with UART (first graph) and I2C (second graph), and for how many bits. }
    \label{fig:allcompatibility}    
\end{figure}

Polyglot frames are exploited by \system both in reading and writing operations. 
While writing a polyglot frame leaves room for the attacker to edit the frame to make it compliant with the chosen protocol (i.e., SPI, UART, I2C), the same editing clearly can not be done while reading. To study such compatibility, e.g., for reading other \acp{ECU} data, we evaluate the compatibility of polyglot frames between real-world \ac{CAN} traffic obtained from the ReCAN dataset~\cite{zago2020recan} and constrained protocols~\footnote{SPI is not considered in this evaluation since it does not add any requirements regarding the sequence of transmitted bits} (i.e., UART, I2C).

Figure~\ref{fig:allcompatibility} represents how many of the 260k CAN packets in the dataset under analysis are compatible with UART and I2C in terms of the length of consecutive compatible bits from the beginning of the CAN packet. UART, even considering its constraints, has no issues in reading the complete CAN ID and is, therefore, a valid reading technique for many link-layer attacks in case of a pin conflict between it and the CAN RX line. On the other side, I2C, which is capable of reading only the 14.3\% of CAN IDs, is less viable to read information from the bus. It is important to note, however, that these numbers are meaningful only in case the attacker is using these peripherals to handle or read fixed CAN data. 

In the writing scenario, where the attacker has more control over the message to send, the feasibility of sending full CAN frames, especially with UART, significantly increases.
\section{The \system Framework}
\label{sec:framework}
As discussed in Sections~\ref{sec:approach}~and~\ref{sec:experiments}, exploiting polyglot frames to allow access to link-layer of \ac{CAN} bus without specialized hardware is the core concept behind \systemns. In particular, link-layer access enables the implementations of attacks, and more importantly, the implementation of defence mechanisms such as CopyCAN~\cite{longari2019copycan}.
This paper focuses on the feasibility of this intuition on the most commonly found peripherals (SPI, UART, I2C, and ADC), even as the concept can be extended to many others. On top of this, we tested and implemented \system on some specific platforms. The implementation on others may require adaptations. 

For these reasons we designed the \system framework, publicly available online\footref{url_release}, which is designed to group together all the techniques presented in this paper under a common interface. 
The framework already contains all the code necessary to reproduce the attacks and run our experiments on our chosen platforms.
The final goal of the framework is that of providing an extensible, cross-platform environment created by the collective, that can be extended by researchers when using \system on novel platforms, while hiding platform specific details such as registers location and peripheral settings. To achieve this goal, the framework is logically split into three different layers:

\mypar{Public interface Layer} The Public Interface Layer is in charge
of defining a unified interface for reading and writing bits on the
CAN bus. This interface is meant to be the main point of contact be-
tween user code implementing a specific attack and the underlying
technique used to mount the attack. More specifically, the reading
and writing primitives are provided by two different interfaces: the
Sender interface and the Receiver interface. This enables the user to
use one peripheral for sending and a different peripheral for receiv-
ing bits on the bus, which means that it is not necessary to have a
full conflict between a peripheral and the CAN controller in order
to use its related technique, but also partially overlapping periph-
erals are allowed.

\mypar{Techniques Layer} Techniques are divided into Senders and Re-
ceivers, each of which implements one of the two public interfaces. Since we do not want to rewrite each technique for
each possible platform, the Platform Layer provides a set of abstract
peripherals that define a minimal interface to interact with each
peripheral. Techniques can use these abstract peripherals to ensure
compatibility with all the supported hardware, which completely
decouples techniques development from platform-specific code.

\mypar{Platform Layer} The Platform Layer provides all the code
related to the interaction with the hardware. Here, we define some of
the most common peripherals, such as I2C, SPI, and GPIO. Each ab-
stract peripheral defines an abstract structure that is implemented
by platform-specific code, through which it can be identified and
passed to the related functions. The abstract peripherals are then implemented in the Platform code. Platforms are the final targets that will be executing the code
(typically microcontrollers). Different vendors provide different
functionalities for each of their microcontrollers, so each platform
might have a slightly different implementation for each peripheral’s
functionality.

\subsection{Comparison with Exisiting Solutions}

Table~\ref{tab:tools-comparison} compares the \system framework with existing solutions that perform \ac{CAN} data-link layer attacks~\cite{CANT,CANhack,Kulandaivel2021cannon}. 
As discussed in Section~\ref{ssec:bit-injection-techniques}, both CANT and CANhack rely on bitbanging, which makes them impractical on low-end microcontrollers.  In addition, the CANT tool is not designed to be cross-platform, while CANhack provides partial portability, although being dependent on the presence of a MicroPython environment. 
Finally, CANnon, which can be considered the state of the art for performing data-link layer attacks, can be theoretically deployed on any microcontroller with an on-chip CAN controller, but it relies on periodic messages and does not provide reading primitives.
On the contrary, the \system framework provides both reading and writing primitives allowing deployment on any microcontroller independently from the periodicity of the messages. The only requirement is the presence of a pin conflict, which is however common in ECUs, as shown in Table~\ref{tab:micro-conflicts}.

\begin{table}
    \centering
    \caption{A comparison of the techniques that can be used to perform data link layer attacks on the \ac{CAN} bus}
    \label{tab:tools-comparison}
    \resizebox{\linewidth}{!}{
    \begin{tabular}{c|ccccc}
         & \textbf{Cross-} & \multirow{2}{*}{\textbf{Read}} &  \multirow{2}{*}{\textbf{Write}} & \textbf{Low-end} & \textbf{Non-periodic} \\
         & \textbf{Platform} & &  & \textbf{MCUs} & \textbf{Messages} 
         \\
         \hline
         \textbf{CANnon}~\cite{Kulandaivel2021cannon} & \checkmark &  & \checkmark & \checkmark &  \\
         \textbf{CANT}~\cite{CANT} &   & \checkmark & \checkmark &  &  \checkmark \\
         \textbf{CANhack}~\cite{CANhack} & $\sim$ & \checkmark & \checkmark &  & \checkmark \\
         \textbf{\systemns} & \checkmark & \checkmark & \checkmark & \checkmark & \checkmark \\
    \end{tabular}
    }
\end{table}
\section{Detection and Prevention of \system}
\label{sec:discussion}

The vast majority of the current literature on intrusion detection for \ac{CAN} focuses on application layer countermeasures~\cite{al2019intrusion}. These software countermeasures interface with the \ac{CAN} controller to receive only the ID and the payload of a correctly received packet. Therefore, they can only process such information. As mentioned in Section~\ref{sec:background}, the \ac{CAN} controller does not forward to the microcontroller any information regarding errors and discarded packets, nor a bit-by-bit view of the bus. For this reason, these \acp{IDS} cannot detect attacks thoughtfully implemented through \system or  exploiting  link-layer attacks in the literature~\cite{palanca2017stealth,Kulandaivel2021cannon,cho2016error,bloom2021weepingcan,yue2021cancloack,murvay2017dos}.

Defenses at the data-link layer are much less common on \ac{CAN} due to the lack of perceived threats. However, it is important to mention CopyCAN~\cite{longari2019copycan} and secure \ac{CAN} transceivers like the NXP TJA115x~\cite{elend2017cyber,nxpstinger}. The first implements a technique to calculate the transmit error counters of \acp{ECU} by reading bus events, therefore detecting an attacker only when it sends spoofed frames after forcing the victim into a bus-off state. The second is a secure \ac{CAN} transceiver that filters incoming and outgoing \ac{CAN} frames by their ID. Moreover, these transceivers can act as a tamper protection mechanism and invalidate messages on the bus in case of spoofing. Both are theoretically viable solutions to limit the capabilities of an attacker that exploits \systemns. Furthermore, one of the most significant drawbacks of CopyCAN~\cite{longari2019copycan} is the requirement of slow bitbanging techniques to read the bus. 
It is worth mentioning that our approach significantly lowers computation requirements for link-layer protection mechanisms, thus enabling CopyCAN implementations on low-end hardware. 

Finally, countermeasures have also been designed at the physical layer. Multiple works~\cite{cho2017viden,choi2018voltageids} demonstrate the use of voltage fingerprinting of \acp{ECU} to recognize spoofing. 
These techniques are, currently, only partially effective in mitigating \systemns. In fact, since they are designed to recognize spoofing of complete frames, they do not detect shorter injections, which is the only requirement for many link-layer attacks.

\mypar{\system aware security} On top of the already existing security solutions, which can partially limit \system capabilities, the most effective countermeasure against \system is avoiding pin conflicts with \ac{CAN} peripherals. We envision this to be obtainable by either choosing an \ac{ECU} microcontroller with knowledge of \system (i.e., choosing hardware that lacks pin conflicts) or by designing the microcontroller itself without conflicts between \ac{CAN} pins and other peripherals. 
Similarly, employing an external \ac{CAN} controller removes the exploitability of pin conflicts, hence securing the platform from \systemns.

\section{Conclusions}
\label{sec:conclusions}

In this paper, we presented \systemns, a novel approach that exploits polyglot frames and pin conflicts to perform data-link layer attacks against CAN, making use of different peripherals already present on the microcontroller. \system enables an attacker to exploit known vulnerabilities of the \ac{CAN} protocol to remotely implement read and write attacks without any assumption on the periodicity of the transmitted messages.  

We experimentally validated \system by studying its feasibility in exploiting existing peripherals protocols, its effectiveness and efficiency in deploying existing attacks, and its compatibility with both low- and high-end microcontrollers on real CAN traffic.  
First, we demonstrated the feasibility of \system on some of the most common peripherals found on standard platforms (i.e., SPI, UART, I2C, and ADC). 
Then, we verified the effectiveness and efficiency of our intuitions by implementing full \ac{CAN} communication between one standard \ac{CAN} node and one where \system is implemented, showing that our techniques heavily reduce the computational requirements for link-layer attacks with respect to standard bitbanging techniques, enabling such attacks from remote on low-end microcontrollers. Moreover, we proved the effectiveness of \system even with partial conflicting peripherals by implementing a targeted denial of service attack that uses SPI and I2C to respectively read and write on the bus.
Finally, we evaluate the compatibility of polyglot frames between real-world CAN traffic and the UART and I2C protocols. 
We provide the community the \system framework to enable the implementation of our approach on different platforms and peripherals through an easy-to-use and expandable interface. The framework allows the exploitation of pin conflicts and polyglot signals in order to build CAN countermeasures and attacks at the data-link layer.
Future works will focus on extending the framework with new platforms and new peripherals, evaluating the feasibility of applying our intuitions on other protocols aside \ac{CAN}, and in the design of data-link layer countermeasures to detect \ac{CAN} attacks through the use of \systemns.

\bibliographystyle{ACM-Reference-Format}
\bibliography{bibliography}

\end{document}